\documentclass[12pt]{article}

\usepackage{times}
\usepackage{graphicx}
\usepackage{amssymb}
\usepackage{amsthm}
\usepackage{amsmath}
\usepackage{dsfont}
\usepackage{bm}
\usepackage{mathrsfs}
\usepackage{bbold}
\usepackage{color}
\usepackage{hyperref}
\usepackage{booktabs} 
\usepackage[table]{xcolor}
\usepackage{array}
\usepackage{caption}

\DeclareMathOperator{\Tr}{Tr}

\begin{document}

\title{Einstein's Equations in Electromagnetic Media}

\author{Eren Erberk Erkul$^{|\dagger\rangle}$ and Ulf Leonhardt$^{|\star\rangle}$\\[0.8em]
\makebox[\textwidth][c]{$^{|\dagger\rangle}$ Department of Physics \& Department of Electrical and Electronics Engineering,}\\
\makebox[\textwidth][c]{Middle East Technical University, Ankara 06800, Turkey}\\[0.8em]
\makebox[\textwidth][c]{$^{|\star\rangle}$ Department of Physics of Complex Systems,}\\
\makebox[\textwidth][c]{Weizmann Institute of Science, Rehovot 7610001, Israel}}

\date{\today}

\maketitle

\begin{abstract}

In this paper, we extend Plebanski's mapping to encode the Einstein equations in ADM form within a bianisotropic electromagnetic medium. We realise this by translating the ADM constraints and evolution equations into dynamical conditions on the medium's constitutive parameters. These transformed equations are then linearised in vacuum to derive gravitational-wave analogues as perturbations of the optical medium.

\end{abstract}

In 1960, Jerzy Plebanski\cite{Plebanski1960}, building on earlier work\cite{Gordon1923,Tamm1924,Skrotskii1957}, and revisited in various forms in numerous follow-up studies\cite{LandauLifshitz1971,deFelice1971,ThorneMacdonald1982}, discovered the mapping,
\begin{equation}
\varepsilon_{ij}= \mu_{ij}= -\frac{\sqrt{-g}\,g_{ij}}{g_{00}}\,,\quad
g_i=\frac{g_{0i}}{g_{00}}\,,\quad
\bm{g}=g_i\,\hat{\bm e}^{\,i},
\label{eq:Plebanski_map}
\end{equation}
that expresses Maxwell's equations in a curved-spacetime metric $g_{\alpha\beta}$ as Maxwell's equations in a flat background but filled with a bianisotropic dielectric medium. This ``material of space'' obeys the constitutive relations
\begin{subequations}\label{eq:constitutive}
\begin{align}
\bm{\mathcal{D}} &= \varepsilon_{0}\varepsilon\,\bm{E}
                   +\frac{\bm{g}}{c}\times\bm{H},
                   \label{eq:constitutiveD}\\
\bm{B}           &= \mu_{0}\mu\,\bm{H}
                   -\frac{\bm{g}}{c}\times\bm{E}.
                   \label{eq:constitutiveB}
\end{align}

\end{subequations}
in SI units, where $\bm{\mathcal{D}}$ is the electric-displacement field, $\bm{E}$ the electric field, $\bm{H}$ the magnetic-field intensity, $\bm{B}$ the magnetic flux density, $\varepsilon$ and $\mu$ are the relative permittivity and permeability tensors, and $\bm{g}$ is the magneto-electric vector with spatial components $g_i$.

Several decades later, the field of transformation optics developed, in which coordinate transformations were shown to be a powerful design tool.\cite{Leonhardt2006,Pendry2006,LeonhardtPhilbin2009,Ozgun2010,WardPendry1996,Chen2010}

However, Plebanski's work, as well as the various analogue-gravity frameworks in optical media that subsequently emerged\cite{Bekenstein2015,Genov2009,NovelloVisserVolovik2002,Leonhardt2008,Barcelo2011}, did not assume that $g_{\alpha\beta}$ solves Einstein's equations in vacuum; instead, $g_{\alpha\beta}$ was taken as the background geometry in which Maxwell's equations are written.

In the present paper, we move beyond the purely kinematic aspects\cite{LeonhardtPhilbin2006}  of this mapping and explore the conditions necessary to encode the dynamical structure of Einstein's equations within dielectric media. We thus require that the transformed optical metric also satisfy the gravitational constraints. This step is motivated by a conceptual similarity between transformation optics and the ADM formalism.\cite{SchusterVisser2024}

Both in the ADM formalism and transformation optics, the object described by the metric is a three-dimensional slice representing a spatial hypersurface. However, in the ADM formalism, one explicitly evolves these spatial slices that are called foliations through time, whereas in transformation optics the analysis is typically confined to a single effective optical slice. Hence the idea is to extend Plebanski's mapping \eqref{eq:Plebanski_map} to the whole ADM system.

It is important to note that this is an analogue encoding at the level of the \((3{+}1)\) field equations; we do not claim a canonical equivalence between the fundamental Hamiltonian phase spaces of general relativity and Maxwell theory. Therefore, the mapped ADM equations should be read as design constraints on constitutive parameters that vary in space and time, and may have to be implemented through active or externally controlled media, rather than as a microphysical derivation of a material whose internal dynamics enforces these conditions.

\textbf{Electromagnetic ADM (3{+}1) System.} The Arnowitt-Deser-Misner (ADM)\cite{ADM1962}, also known as the Hamiltonian formalism, is an equivalent formulation of general relativity.\cite{Misner1973} Unlike the usual covariant picture, this formulation reduces the spacetime manifold to a \emph{trivial} fibre \cite{Gourgoulhon2007,Nakahara2003}:
\begin{equation}
\label{eq:trivialbundle}
\pi:\; E \equiv E_{1}\times E_{3}\;\longrightarrow\; E_{1},
\qquad
\pi(t,x)=t .
\end{equation}
Here $E_{1}\simeq\mathbb{R}$ is the base (global Cauchy) while $E_{3}$ denotes the leaves of the foliation that plays the role of the fibre and is homeomorphic to a Cauchy surface.
\footnote{
In GR, the collection of fibres through base space can be viewed as the collection of inertial frames, where a suitable choice of coordinates (an orientation of foliation) can always flatten each fibre into an inertial one. The key difference between the \emph{trivial} bundles of GR and those of Newton--Cartan theory is an extra degree of freedom that allows the spatial foliation to twist and curl. Hence, Newtonian foliations are absolute entities; no gravitational waves arise there, lacking radiative TT part of the curvature. Relatedly, both Newtonian and Einsteinian bundles differ from the Galilean bundle in that the extra freedom now resides in the fibres themselves.\cite{Penrose2004,Trautman1965}}

\begin{figure}[htbp]
  \centering
  \includegraphics[width=\linewidth,height=0.3\textheight,keepaspectratio]{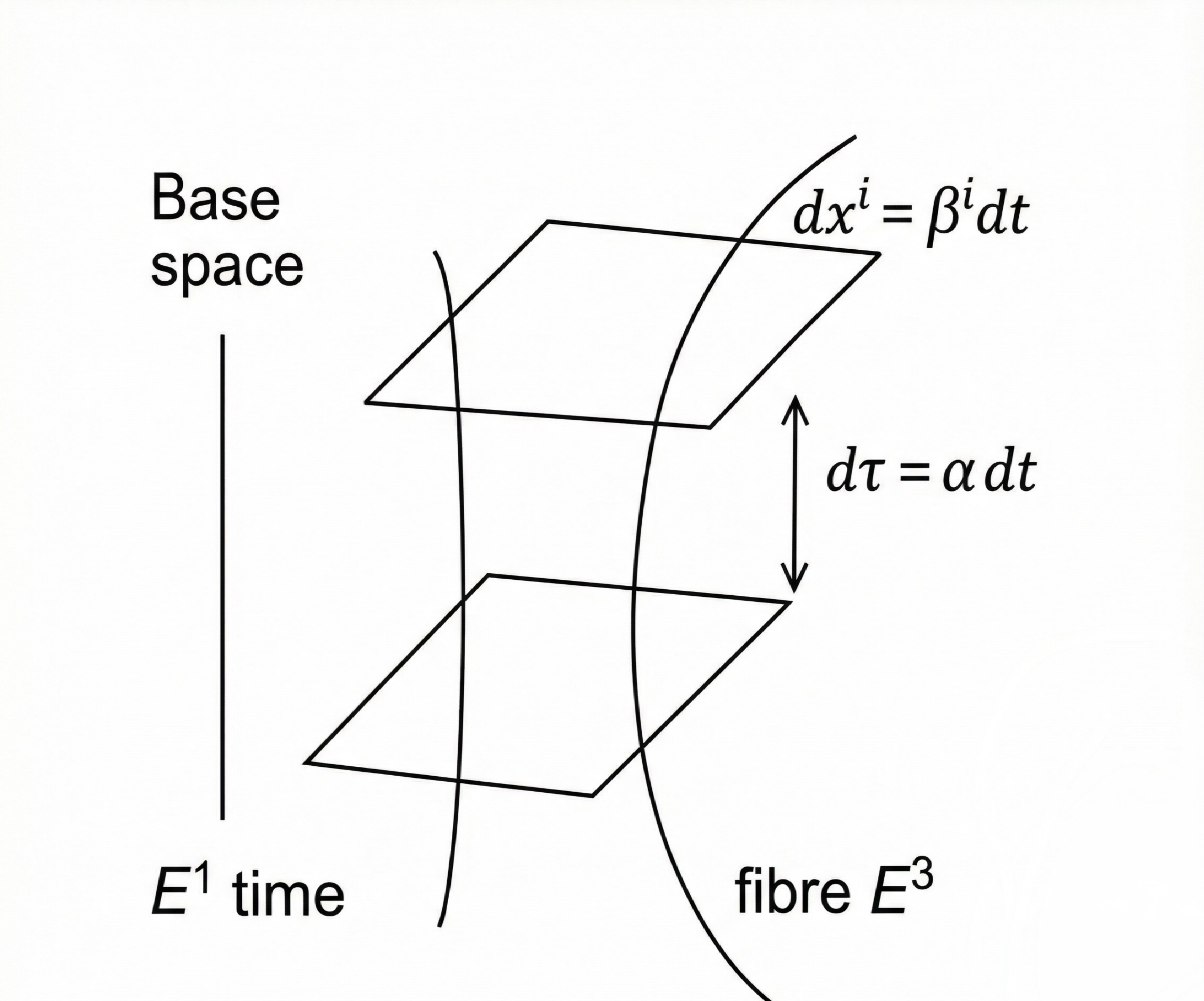}
  \caption{Visualisation of ADM foliation}
  \label{fig:ADM_bundle}
\end{figure}

The line element in ADM coordinates adapted to the slicing defined by \(t=\text{constant}\) reads
\begin{equation}
ds^{2}=g_{ab}\,dx^{a}dx^{b}
      =-\alpha^{2}dt^{2}
      +\gamma_{ij}\bigl(dx^{i}+\beta^{i}dt\bigr)\bigl(dx^{j}+\beta^{j}dt\bigr).
\label{eq:line}
\end{equation}
Equivalently, the metric tensor in matrix form reads
\begin{align}
g_{ab} &=
\begin{pmatrix}
-\alpha^{2} + \beta_{i}\beta^{i} & \beta_{i} \\
\beta_{j} & \gamma_{ij}
\end{pmatrix},
\\[6pt]
g^{ab} &=
\begin{pmatrix}
-\alpha^{-2} & \alpha^{-2}\beta^{i} \\
\alpha^{-2}\beta^{j} & \gamma^{ij} - \alpha^{-2}\beta^{i}\beta^{j}
\end{pmatrix}.
\end{align}
Here, $\gamma_{ij}$ is the spatial metric on the $t=\text{constant}$ slices. 
The dynamism of this system is encoded in the \emph{lapse} $\alpha$ and \emph{shift} $\beta^{i}$, both one-forms. While the net evolution vector, called the \emph{time vector} $t^{\mu}$, drags a point through the full \((3{+}1)\)-dimensional spacetime, it combines both effects:
\begin{equation}
  \label{eq:timevector}
  t^{\mu} = \alpha\,n^{\mu} + \beta^{\mu},
\end{equation}
with $n^{\mu}$ the future-pointing unit normal to each $\Sigma_{t}$ ($n^{\mu}n_{\mu}=-1,\;n^{\mu}\gamma_{\mu\nu}=0$).

For a detailed literature of the \(3{+}1\) decomposition, see, e.g.\ \cite{Baumgarte2010,Rezzolla2013,NovelloVisserVolovik2002}.

\textbf{Shift as Medium Drag.} Shift vectors gauge how points move within a spatial slice when we advance the coordinate time:
\begin{equation}
  \label{eq:shift-diff}
  dx^{i} = \beta^{i}\,dt
  \qquad (\text{on } \Sigma_{t}).
\end{equation}
Extending Plebanski's mapping \eqref{eq:Plebanski_map} and substituting the ADM decomposition
$g_{0i}=\gamma_{ij}\beta^{j}$:
\begin{equation}
g_i = \frac{\gamma_{ij}\beta^{j}}{g_{00}},
\end{equation}
where \(g_{i}\) (one spatial index) is the medium's
\emph{magneto-electric} vector, not a component of the metric. Raising the spatial index with $\gamma^{ij}$ gives
\begin{equation}
\gamma^{ij}g_i = \frac{\beta^{j}}{g_{00}}.
\end{equation}
So, the same geometric information in \(g_{i}\) is encoded in the shift vector:

\begin{equation}
  \beta^{i}= g_{00}\,\gamma^{ij}g_{j}.
  \label{eq:beta-g-relation}
\end{equation}

Hence a nonzero $\beta^{i}$ is realised optically as a cross-coupling term ${(\mathbf g/c)\!\times\!\mathbf H}$ in the constitutive relations, which is exactly what one obtains for a dielectric moving with a physical three-velocity proportional to $\mathbf g$.  Equivalently,
$\beta^{i}$ can be regarded as an \emph{effective} medium velocity
$\mathbf v_{\text{eff}}\propto\mathbf g$, which, through Eq.~\eqref{eq:beta-g-relation},
induces the very Lorentz-type mixing of $\mathbf E$ and $\mathbf B$ that a real boost generates
in free space.

\newpage

\textbf{Lapse as Conformal Factor.} Lapse $\alpha$ is a one-form that measures proper time between adjacent slices:
\begin{equation}
  d\tau=\alpha\,dt .
\end{equation}

Since the source-free Maxwell equations in four dimensions are invariant under local conformal rescalings
\(g_{\mu\nu}\rightarrow\tilde g_{\mu\nu}=\Omega^{2}g_{\mu\nu}\),
we can factor out a positive scalar $\Omega^{2}(x)$ from the metric\cite{LeonhardtPhilbin2009},
\begin{equation}
\label{eq:conformal}
  g_{\mu\nu}=\Omega^{2}\,\widehat g_{\mu\nu}.
\end{equation}
A convenient choice, equivalent to taking the  $t=\text{constant}$  slicing, is to set $\widehat g_{00}=-1$ in the ADM line element (Eq.~\eqref{eq:line}), which fixes
\begin{equation}
  \label{eq:Omega-general-der}  
  \Omega^{2}(x)=-g_{00}
               =\alpha^{2}-\gamma_{ij}\beta^{i}\beta^{j},
\end{equation}

Taking the positive square root gives
\begin{equation}
  \Omega(x)=\sqrt{\alpha^{2}-\gamma_{ij}\beta^{i}\beta^{j}} .
  \label{eq:Omega-general}
\end{equation}

When the shift vanishes ($\beta^{i}=0$), this reduces to $\Omega=\alpha(t,\mathbf x)$, matching the standard result in transformation optics: changing the lapse is merely a conformal rescaling of the metric.

The index-lowered vacuum expressions for the Plebanski constitutive tensors (\ref{eq:Plebanski_map}) with identification (\ref{eq:Omega-general-der}) read
\begin{equation}
  \varepsilon_{ij}=\frac{\sqrt{\gamma}}{\alpha}\,\gamma_{ij},
  \qquad
  \mu_{ij}        =\frac{\sqrt{\gamma}}{\alpha}\,\gamma_{ij}.
\end{equation}
Under the conformal rescaling
(\ref{eq:conformal}) the ADM variables transform as
\begin{equation}
  \tilde\alpha=\Omega\,\alpha,\qquad
  \tilde\gamma_{ij}=\Omega^{2}\gamma_{ij},\qquad
  \sqrt{\tilde\gamma}=\Omega^{3}\sqrt{\gamma}.
\end{equation}
Substituting these into the definition of $\varepsilon_{ij}$ gives
\begin{equation}
  \tilde\varepsilon_{ij}
    =\frac{\sqrt{\tilde\gamma}}{\tilde\alpha}\,
      \tilde\gamma_{ij}
    =\frac{\Omega^{3}\sqrt{\gamma}}{\Omega\alpha}\,
      (\Omega^{2}\gamma_{ij})
    =\Omega^{4}\varepsilon_{ij},
\end{equation}
and the same calculation yields
\begin{equation}
  \tilde\mu_{ij}=\Omega^{4}\mu_{ij}.
\end{equation}

Hence, under the same rescaling the Plebanski constitutive tensors acquire four powers of the conformal factor,
\begin{equation}
  \tilde\varepsilon_{ij}=\Omega^{4}\varepsilon_{ij},
  \qquad
  \tilde\mu_{ij}=\Omega^{4}\mu_{ij},
\end{equation}
 As expected, the source-free Maxwell equations are therefore insensitive to the overall scale $\Omega$.

 \clearpage

{It is worth noting that conformal transformations play a different role in lower-dimensional systems. In two spatial dimensions, every metric is locally conformally flat, so the Plebanski map can be satisfied with an isotropic medium,
\begin{equation}
\varepsilon_{ij}=\mu_{ij}=\Omega^{2}\delta_{ij}
\qquad i,j=1,2.
\end{equation}
making conformal methods a powerful design tool in transformation optics~\cite{Leonhardt2006}. By contrast, in three spatial dimensions local conformal flatness is dictated by the Cotton tensor, whereas in the full \((3{+}1)\)-dimensional spacetime conformal curvature is characterised by the Weyl tensor. In this setting, conformal rescalings do not generate an arbitrary spatial metric and are therefore not a viable general design tool. Hence, in the present work we analyse the conformal invariance of the source-free Maxwell equations rather than conformal flatness of the metric itself. This amounts to a gauge redundancy in the Plebanski map, which is then reduced once the ADM dynamical conditions are imposed.
}

The decomposition remains valid provided $g_{00}<0$, i.e.\ $\alpha^{2}>\gamma_{ij}\beta^{i}\beta^{j}$, a condition automatically satisfied for any spacelike foliation.

\textbf{Electromagnetic ADM Constraints and Dynamics} The \(3{+}1\) (ADM) split of spacetime entails two constraint equations\cite{ADM1962} that determine the intrinsic structure of each hypersurface, namely the momentum and Hamiltonian constraints.
\begin{subequations}\label{eq:ADM_constraints}
\begin{align}
{}^{(3)}R+K^{2}-K_{ij}K^{ij} &= 16\pi G\,\rho , \label{eq:ADM_H}\\
D_{j}\!\bigl(K^{ij}-\gamma^{ij}K\bigr) &= 8\pi G\,j^{\,i}, \label{eq:ADM_M}
\end{align}
\end{subequations}

\begin{figure}[htbp]
  \captionsetup{skip=0pt, justification=centering}

  \makebox[0.9\linewidth][r]{%
    \includegraphics[width=0.8\linewidth,height=0.3\textheight,keepaspectratio]{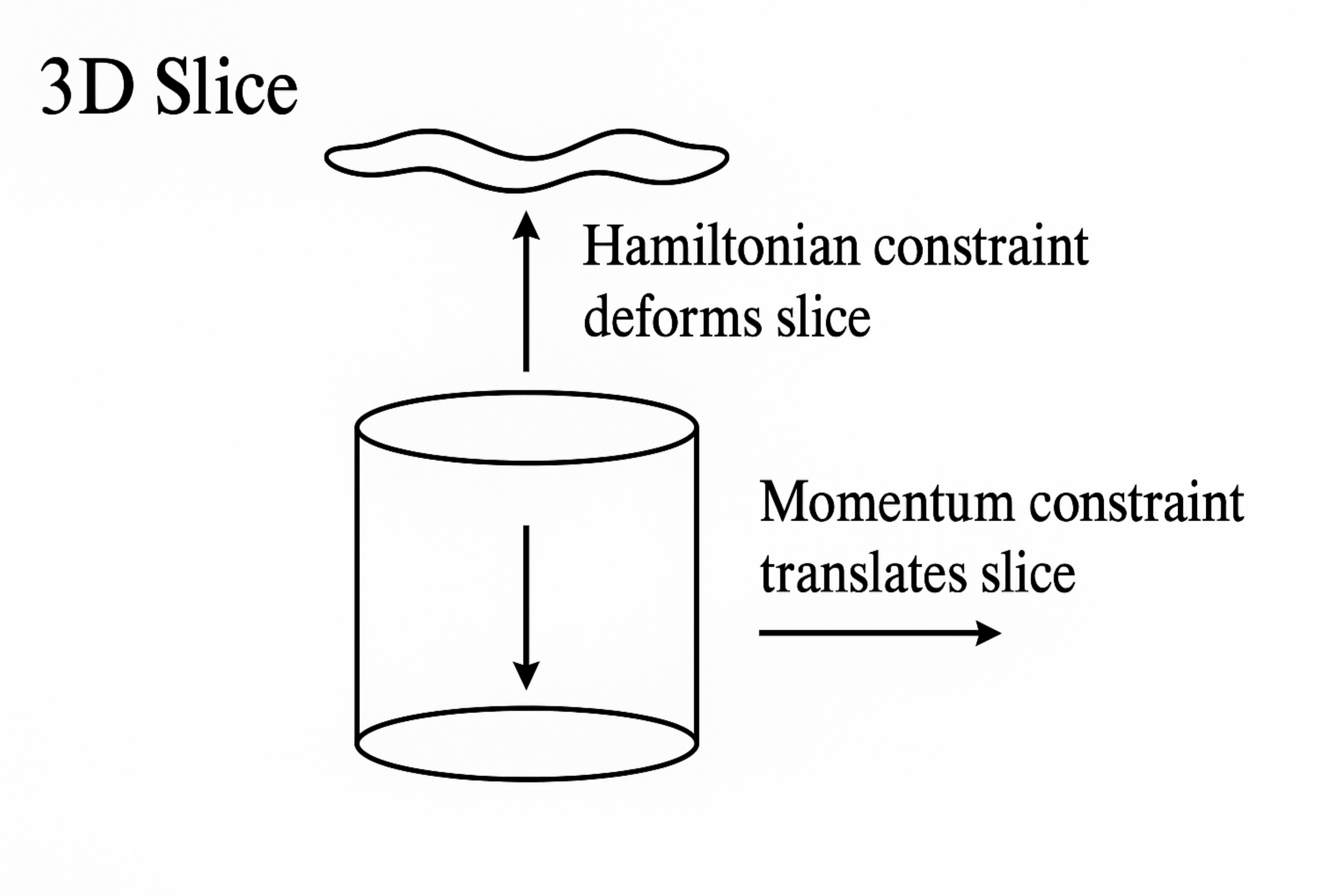}
  }
  \caption{Visualisation of ADM constraint equations.}
  \label{fig:ADM_constraints}
\end{figure}

\vspace{1em}

\noindent
where $\rho$ is the energy density, $j^{\,i}$ the momentum density,
$K_{ij}$ the extrinsic curvature, ${}^{(3)}\!R_{ij}$ (and its trace
${}^{(3)}\!R$) the intrinsic curvature of the spatial hypersurface, and
$D_i$ the covariant derivative compatible with $\gamma_{ij}$.

\clearpage

Together with the evolution \cite{ADM1962}, which includes the dynamic entities lapse and shift,

\begin{subequations}\label{eq:ADM_evolution}
  \begin{align}
    \partial_{t}\gamma_{ij} &=
      -2\alpha K_{ij}+D_{i}\beta_{j}+D_{j}\beta_{i},
      \label{eq:ADM_E1}\\[2pt]
    \partial_{t}K_{ij} &=
      -D_{i}D_{j}\alpha
      +\alpha\!\bigl({}^{(3)}R_{ij}+K\,K_{ij}-2K_{ik}K^{k}{}_{j}\bigr) \notag\\
    &\quad
      -8\pi G\,\alpha\!\Bigl(
        S_{ij}-\tfrac12\gamma_{ij}\,(\rho-S)
      \Bigr) \notag\\
    &\quad
      +\beta^{k}D_{k}K_{ij}
      -K_{kj}D_{i}\beta^{k}
      -K_{ik}D_{j}\beta^{k},
      \label{eq:ADM_E2}
  \end{align}
\end{subequations}

where $S^{ij}$ is the spatial electromagnetic stress tensor, \cite{Gourgoulhon2007}
\begin{equation}
S^{ij}=E^{i}\mathcal{D}^{j}+H^{i}B^{j}
       -\tfrac12\gamma^{ij}\bigl(\bm{E}\!\cdot\!\bm{\mathcal{D}}
       +\bm{B}\!\cdot\!\bm{H}\bigr).
\label{eq:S_EM}
\end{equation}
Since we confine ourselves to electromagnetic fields and Maxwell stress-energy is traceless \cite{LandauLifshitz1971}, 
\(S\equiv\gamma_{ij}S^{ij}=\rho c^{2}\), the \(\gamma_{ij}(\rho-S)\) term vanishes and the evolution equation retains only the electromagnetic back-reaction, which introduces non-linearities into the sources discussed in the next section

Solving \eqref{eq:ADM_E1} for the extrinsic curvature gives
\begin{equation}
  K_{ij}
  = -\frac{1}{2\alpha}\bigl(\partial_t\gamma_{ij}-D_i\beta_j-D_j\beta_i\bigr).
  \label{eq:K_ADM_general}
\end{equation}

Reorganised, the Plebanski map \eqref{eq:Plebanski_map} reads,

\begin{equation}
  \gamma_{ij}= -\frac{g_{00}}{\sqrt{-g}}\;\varepsilon_{ij},
  \qquad
  \beta_i=g_{0i}=g_{00}\,g_i ,
  \label{eq:identifications}
\end{equation}
with \(\sqrt{-g}=\alpha\sqrt{\gamma}\).  Substitution into
\eqref{eq:K_ADM_general} yields
\begin{equation}
  K_{ij}
  = -\frac{1}{2\alpha}\Bigl[
    \partial_t\!\Bigl(-\tfrac{g_{00}}{\sqrt{-g}}\varepsilon_{ij}\Bigr)
    - D_i\!\bigl(g_{00}g_j\bigr)
    - D_j\!\bigl(g_{00}g_i\bigr)
  \Bigr],
  \label{eq:K_optical_general}
\end{equation}
where the first term captures explicit temporal modulation and the
remaining covariant derivatives describe stationary ``drag'' induced by
the gauge fields.  Both contributions appear quadratically in
\eqref{eq:ADM_H}.

Because \(\beta^{m}=g_{00}g^{m}\),
\(g_{00}=-\alpha^{2}+\gamma_{mn}\beta^{m}\beta^{n}\) becomes a quadratic
equation whose negative-root solution is
\begin{equation}
  g_{00}= \frac{1-\sqrt{1+4\alpha^{2}S}}{2S},
  \qquad
  S\equiv\gamma^{mn}g_m g_n .
  \label{eq:g00_solution}
\end{equation}
Equations \eqref{eq:K_optical_general} and \eqref{eq:g00_solution}
express \(K_{ij}\) entirely in terms of
\(\{\varepsilon_{ij},g_i,\alpha,\beta^{i}\}\).

The intrinsic geometry of each optical slice is encoded in
\(\gamma_{ij}(t,\mathbf x)\equiv g_{ij}\), from which one builds
\({}^{(3)}R_{ij}\) and its scalar\cite{Wald1984}
\begin{equation}
  {}^{(3)}R=\gamma^{ij}\,{}^{(3)}R_{ij}.
  \label{eq:R3-def}
\end{equation}

\clearpage

Under a conformal rescale (\ref{eq:conformal}),
\begin{equation}
  {}^{(3)}\!\widetilde R
  =\Omega^{-2}\!\Bigl(
      {}^{(3)}\!R
      -4\,\Delta\!\ln\Omega
      -2\,\gamma^{ij}\partial_{i}\ln\Omega\,\partial_{j}\ln\Omega
    \Bigr),
  \label{eq:R3-conformal}
\end{equation}
with \(\Delta=\gamma^{ij}D_{i}D_{j}\).  Choosing
\(\Omega^{2}= -g_{00}= \alpha^{2}-\gamma_{kl}\beta^{k}\beta^{l}\) gives
\begin{equation}
\label{eq:R3-full}%
\begin{aligned}
{}^{(3)}\!R[\varepsilon;\alpha,\beta]
  &=\frac{
        {}^{(3)}\!R[\varepsilon]
        -4\,\Delta\!\ln\alpha_{\text{eff}}
        -2\,\gamma^{ij}\partial_{i}\ln\alpha_{\text{eff}}
          \,\partial_{j}\ln\alpha_{\text{eff}}}
        {\alpha^{2}-\gamma_{kl}\beta^{k}\beta^{l}}
\\[6pt]
\alpha_{\text{eff}}
  &\equiv
    \sqrt{\alpha^{2}-\gamma_{kl}\beta^{k}\beta^{l}} .
\end{aligned}
\end{equation}

The numerator depends only on the spatial permittivity, whereas lapse
and shift enter through the conformal prefactor and its gradients. 

Henceforth, every appearance of \(\gamma_{ij},\,K_{ij},\,{}^{(3)}R_{ij}\)
or \(D_i\) refers to the optical metric
\(\gamma_{ij}[\varepsilon,\mu]\) of~(\ref{eq:optical_metric}).

\textbf{Electromagnetic ADM Sources.}

Given mass sources and currents, the ADM constraints (\ref{eq:ADM_constraints}) and evolution (\ref{eq:ADM_evolution}) encode Einstein's equations.  
Using the identifications (\ref{eq:identifications}), the entire system can be written in the language of the effective dielectric. To complete the analogue system, we take both the mass density $\rho$ and the momentum density $\bm{j}$ to be provided solely by the electromagnetic field. This would describe the self-gravity of light. Note that this is not an entirely artificial situation. For example, the early Universe was radiation-dominated; black-hole magnetospheres can be modelled electromagnetically \cite{Komissarov2004}; and certain black-hole metrics can be generated from colliding electromagnetic waves through the Chandrasekhar--Xanthopoulos duality \cite{Halilsoy2025,Chandrasekhar1986}.

With this choice our source expressions coincide with those obtained by Thorne and Macdonald in their \(3{+}1\) formulation of electrodynamics.\cite{ThorneMacdonald1982,MembraneParadigm1986}

In the notation we already established, the Maxwell's equations in the \(3{+}1\) formalism read:
\begin{subequations}\label{eq:Maxwell_TM}
\begin{align}
D_{i}B^{i} &= 0, \label{eq:Maxwell_TM1}\\[2pt]
\partial_{t}\!\bigl(\sqrt{\gamma}\,B^{i}\bigr)
  &+ \partial_{j}\!\Bigl[\sqrt{\gamma}\bigl(\beta^{j}B^{i}-\beta^{i}B^{j}
           -\alpha\epsilon^{ijk}E_{k}\bigr)\Bigr]=0, \label{eq:Maxwell_TM2}\\[4pt]
D_{i}\mathcal{D}^{i} &= \sqrt{\gamma}\,\rho_{\mathrm e}, \label{eq:Maxwell_TM3}\\[2pt]
\partial_{t}\!\bigl(\sqrt{\gamma}\,\mathcal{D}^{i}\bigr)
  &+ \partial_{j}\!\Bigl[\sqrt{\gamma}\bigl(\beta^{j}\mathcal{D}^{i}-\beta^{i}\mathcal{D}^{j}
           +\alpha\epsilon^{ijk}H_{k}\bigr)\Bigr] \notag\\[2pt]
  &\qquad = -\sqrt{\gamma}\,\alpha\,\mathcal J^{\,i}. \label{eq:Maxwell_TM4}
\end{align}
\end{subequations}

where $\rho_e \equiv n_\mu J^\mu$ is the electric charge density measured by an Eulerian observer, $J^{i}\equiv\gamma^{i}_{\ \mu}J^\mu$ is the coordinate three--current density, $\mathcal{J}^{\,i}\equiv J^{i}+\rho_e\beta^{i}$ is the physical current seen by an Eulerian observer and $\epsilon^{ijk}$ is the Levi-Civita symbol associated with the spatial metric. 

The individual expressions can be understood within our framework when the identifications in (\ref{eq:identifications}) are applied.
\clearpage
Projecting the electromagnetic stress--energy tensor~\cite{Misner1973}, one finds that, in the analogy, the mass-density $\rho$ is identified with the energy density and the current $\mathbf{j}$ with the Poynting vector, both divided by $c^{2}$:
\cite{ThorneMacdonald1982}
\begin{subequations}\label{eq:rho_j_S}
\begin{align}
\rho &= \frac{1}{2c^{2}}\bigl(\bm{\mathcal{D}}\!\cdot\!\bm{E}
           +\bm{B}\!\cdot\!\bm{H}\bigr), \label{eq:rho_EM}\\[4pt]
j^{\,i} &= \frac{1}{c^{2}}\bigl(\bm{E}\times\bm{H}\bigr)^{i}, \label{eq:j_EM}
\end{align}
\end{subequations}

In this way, we describe the nonlinear gravitational dynamics
of light through the dielectric-medium equations; solving
them yields the full dynamics of an optical medium that is the
analogue of a self-gravitating spacetime.

\textbf{Analogue of Gravitational Waves.} Now that we have the machinery at hand, let us consider its implications for the dielectric by examining the linearised regime of general relativity, which is as simple as it gets, but not simpler. Assume $g_{00}=-1$, so the lapse is $\alpha=1$ and the shift is $\beta^{i}=0$ throughout this section.  
For the dielectric tensors we obtain, from Eq.~(\ref{eq:Plebanski_map}),
\begin{equation}
\gamma_{ij}= (\det\varepsilon)^{-1/5}\,\varepsilon_{ij},
\label{eq:optical_metric}
\end{equation}
a form that also arises in many applications of transformation optics \cite{LeonhardtPhilbin2009}.

In the regime of weak gravity,
\begin{equation}
g_{ij} = \delta_{ij} + h_{ij}, \qquad |h_{ij}|\ll 1,
\label{eq:weak}
\end{equation}
At linear order, the term $-8\pi G\,\alpha S_{ij}$ in
Eq.~\eqref{eq:ADM_E2} is $\mathcal{O}(h^{2})$ and is therefore
neglected here.
Evolution equations \eqref{eq:ADM_E1} and \eqref{eq:ADM_E2} reduce to
\begin{subequations}\label{eq:K_evolution}
  \begin{align}
    K_{ij} &= -\tfrac12\,\partial_{t}h_{ij},\label{eq:K_evolution1}\\[2pt]
    \partial_{t}K_{ij} &= {}^{(3)}\!R^{\mathrm{lin}}_{ij}, \label{eq:K_evolution2}
  \end{align}
\end{subequations}
where the linearized Ricci tensor ${}^{(3)}\!R^{\mathrm{lin}}_{ij}$ is given by \cite{Baumgarte2010}
\begin{equation}
{}^{(3)}\!R^{\mathrm{lin}}_{ij}
  = -\tfrac12 \nabla^{2} h_{ij}
    -\tfrac12 \partial_{i}\partial_{j} h
    +\tfrac12 \partial_{(i}\partial^{k} h_{j)k}.
\label{3Riccilin_full}
\end{equation}

Assuming vacuum (\(\rho=\bm{j}=0\)) and the weak-gravity regime \eqref{eq:weak}, the Hamiltonian constraint (\ref{eq:ADM_H}) yields \cite{Gourgoulhon2007}
\begin{equation}
\partial_{j}\partial_{k}h^{jk}-\nabla^{2}h = 0, \qquad h\equiv\delta^{ij}h_{ij}.
\label{Ham}
\end{equation}

The momentum constraint (\ref{eq:ADM_M}) gives \cite{Gourgoulhon2007}
\begin{equation}
\partial_{j}\dot h^{ij}-\partial^{i}\dot h = 0,
\label{Mom}
\end{equation}
with \(\dot h_{ij}\equiv\partial_{t}h_{ij}\) and \(\dot h \equiv \delta^{kl}\dot h_{kl}\).
\clearpage
Under the transverse-traceless (TT) conditions\cite{Misner1973}:
\begin{equation}
\label{eq:TT}
\begin{aligned}
\partial^{i}h^{TT}_{ij} &= 0,\\[2pt]
h^{TT} &= 0,
\end{aligned}
\end{equation}
(\ref{Ham}) and (\ref{Mom}) are trivially satisfied.

Continuing in this gauge, Eq.~(\ref{3Riccilin_full}) reduces to
\begin{equation}
R_{ij}^{\text{lin}}=-\tfrac12\nabla^{2}h_{ij}.
\label{TTRicci}
\end{equation}

Finally, substituting (\ref{eq:K_evolution1}) and (\ref{TTRicci}) into (\ref{eq:K_evolution2}), we isolate the two radiative degrees of freedom and obtain
\begin{equation}
\frac{1}{c^2}\partial_{t}^{2}h^{TT}_{ij}-\nabla^{2}h^{TT}_{ij}=0.
\label{GW}
\end{equation}

We write the permittivity as a small perturbation about the identity:
\begin{equation}
\varepsilon_{ij}=\delta_{ij}+\delta\varepsilon_{ij},
\qquad
|\delta\varepsilon_{ij}|\ll1 .
\end{equation}
Expanding the determinant to first order gives
\begin{equation}
\label{eq:det_expansion}
\begin{aligned}
\det\varepsilon     &= 1+\Tr(\delta\varepsilon)
                       +\mathcal{O}\!\bigl((\delta\varepsilon)^2\bigr),\\[2pt]
(\det\varepsilon)^{-1/5} &= 1-\tfrac15\Tr(\delta\varepsilon)
                       +\mathcal{O}\!\bigl((\delta\varepsilon)^2\bigr).
\end{aligned}
\end{equation}

Inserting these into Eq.~\eqref{eq:optical_metric}, we find
\begin{align}
\gamma_{ij}
  &= (\det\varepsilon)^{-1/5}\,\varepsilon_{ij}                \nonumber\\
  &= \delta_{ij} + \delta\varepsilon_{ij}
     - \tfrac15\delta_{ij}\Tr(\delta\varepsilon)
     + \mathcal{O}\!\bigl((\delta\varepsilon)^2\bigr), \\[2pt]
h_{ij}
  &\equiv \gamma_{ij}-\delta_{ij}
  =      \delta\varepsilon_{ij}
     - \tfrac15\delta_{ij}\Tr(\delta\varepsilon)
.
\label{eq:h_from_deps}
\end{align}

Any symmetric tensor \(X_{ij}\) can be decomposed into trace, longitudinal, and transverse-traceless parts:
\begin{equation}
X_{ij}=X^{\mathrm{TT}}_{ij}+\partial_{(i}V_{j)}
       +\tfrac13\delta_{ij}\Tr(X).
\end{equation}
Applying this projector to Eq.~\eqref{eq:h_from_deps} and using the gauge conditions (\ref{eq:TT}), we identify the dielectric perturbation with the metric wave:
\begin{equation}
h^{\mathrm{TT}}_{ij}=\delta\varepsilon^{\mathrm{TT}}_{ij}\,.
\label{eq:TT_dictionary}
\end{equation}

Substituting into (\ref{GW}) gives
\begin{equation}
\frac{1}{c^2}\partial_{t}^{2}\,\delta\varepsilon^{TT}_{ij}
  -\nabla^{2}\,\delta\varepsilon^{TT}_{ij}=0.
\label{OptGW}
\end{equation}

The vacuum speed $c$ appears in Eq.~\eqref{OptGW} because the linearisation is done with respect to flat space that corresponds to a background vacuum dielectric $\varepsilon_{ij}=\delta_{ij}$. Physically, a medium satisfying Eq.~(\ref{OptGW}) can be implemented using the Kerr effect in an optical medium, similar to the implementation of the optical analogue of the event horizon \cite{Leonhardt2008}. All one has to do is to generate a sinusoidal modulation on a continuous--wave pump beam using, for example, an acousto--optic modulator. For a planar pump wave, the modulation would naturally propagate with Eq.~\eqref{OptGW} exempt that the background medium has an effective refractive index $n_{\mathrm{eff}}\neq 1$ (equivalently, $\varepsilon_{ij}\neq\delta_{ij}$), $c$ is replaced by $c/n_{\mathrm{eff}}$, but the mathematical structure of the wave equation is unchanged.

%\clearpage

Equation~\eqref{OptGW} admits the usual plane-wave solution:

\begin{subequations}\label{planewave}
\begin{align}
\delta\varepsilon^{\mathrm{TT}}_{ij}(\mathbf{x},t)
  &= \operatorname{Re}\!\bigl\{\mathcal{A}_{ij}\,e^{i(\mathbf{k}\cdot\mathbf{x}-\omega t)}\bigr\},\\[4pt]
\intertext{or equivalently}
\delta\mu^{\mathrm{TT}}_{ij}(\mathbf{x},t)
  &= \operatorname{Re}\!\bigl\{\mathcal{A}_{ij}\,e^{i(\mathbf{k}\cdot\mathbf{x}-\omega t)}\bigr\},
\end{align}
\end{subequations}

where the constant, symmetric, traceless tensor \(\mathcal{A}_{ij}\) obeys \(k^{i}\mathcal{A}_{ij}=0\). {For a given propagation direction, the transverse and traceless 
conditions reduce $\mathcal{A}_{ij}$ to two linearly independent components, 
corresponding to the $+$ and $\times$ polarisation states of the gravitational 
wave. Since Eqs.~\eqref{GW} and \eqref{OptGW} are the same PDE, the optical 
correspondence inherits the solutions of the gravitational Cauchy problem 
with the TT conditions holding for both polarisations.} Thus, equation~\eqref{planewave} is a direct analogue of a gravitational wave propagating through the optical medium and can be used as a theoretical model for spatio-temporal perturbations in engineered media. {In particular, the permittivity profile of a spacetime metasurface such as that of Ref.~\cite{Atwater2024} can be electrically tuned to exhibit the wave-like behaviour described by Eq.~\eqref{OptGW}. In this sense, this model serves as a design tool for dynamic media, in the same spirit as Plebanski's mapping~\eqref{eq:Plebanski_map}, which provides a design specification for the static case. The established framework can be extended beyond the linear regime using analytic and numerical methods, and may serve as a modelling tool for elaborate time-varying media such as the photonic time crystals~\cite{Wang2023}.}

If, instead of fixing the shift to zero, we keep a uniform (time-independent) shift $\beta^{i}$, the ordinary time derivative in the wave equation~\eqref{GW} is replaced by the convective derivative\cite{Gourgoulhon2007}
\begin{equation}
  \partial_{t}\;\mapsto\;D_{0}\equiv\partial_{t}-\beta^{k}\partial_{k}\, ,
  \label{eq:D0}
\end{equation}
so the flat d'Alembertian becomes the \emph{boosted} operator\cite{Baumgarte2010}
\begin{equation}
  \Box_{\beta}\;=\;\frac{1}{c^{2}}\,D_{0}^{2}-\nabla^{2},
  \label{eq:BoxBeta}
\end{equation}
i.e.\ the d'Alembertian expressed in Lorentz-boosted coordinates with velocity~$\boldsymbol{\beta}$.

Hence, the plane-wave solution~\eqref{planewave} acquires an extra phase
\begin{equation}
  \exp\!\bigl[\mathrm{i}\,(\mathbf{k}\!\cdot\!\boldsymbol{\beta})\,t\bigr],
  \label{eq:PhaseShift}
\end{equation}
corresponding to the constant Doppler shift introduced by the boost.

When the shift vector $\beta^{i}$ stops being uniform in spacetime, the simple
substitution~(\ref{eq:D0}) no longer leaves the d'Alembertian~(\ref{eq:BoxBeta}) unchanged:
extra geometric terms enter the operator and couple different Fourier modes.
As the Einsteinian system is nonlinear, these additional couplings don't obey
superposition: every harmonic becomes entangled with the background curvature
along with the other modes. Therefore in strong regimes, although the equations
can still be expressed analytically, the gravitational waves cannot be built
from boosted plane waves. One must evolve the analogous \(3{+}1\) system numerically,
which is highly non-trivial.\cite{Pretorius2005,CookTeukolsky1999}

\clearpage

\textbf{Conclusion.}

Building on Plebanski's work, we have shown that electrodynamics in a bianisotropic medium not only has sufficient independent degrees of freedom to map the spacetime metric but also the dynamical structure to encode the ADM field equations of general relativity as conditions on the constitutive parameters (see Table~\ref{tab:correspondence}).

\begin{table}[h]
  \centering
  \caption{ADM--Optical correspondence.}
  \label{tab:correspondence}
  \renewcommand{\arraystretch}{1.2}  
  \setlength{\tabcolsep}{6pt}     
  \rowcolors{2}{gray!5}{white}

  \begin{tabular}{@{}
    >{\centering\arraybackslash}p{0.47\linewidth}
    !{\color{gray!40}\vrule width 0.3pt}
    >{\centering\arraybackslash}p{0.47\linewidth}
    @{}}
    \toprule
    \textbf{ADM Gravity (\(3{+}1\))} & \textbf{Bianisotropic Medium} \\
    \midrule
    Lapse function $\alpha$ & Conformal factor $\Omega$ \\
    Shift vector $\beta^{i}$ & Magneto-electric coupling $\mathbf{g}$ \\
    Spatial metric $\gamma_{ij}$ & Permittivity tensor $\varepsilon_{ij} \equiv \mu_{ij}$ \\
    Extrinsic curvature $K_{ij}$ & Material evolution $\partial_t \varepsilon_{ij}$ \\
    Matter sources $(\rho, j^{i})$ & EM stress-energy \\
    Gravitational waves $h^{\mathrm{TT}}_{ij}$ 
      & Dielectric perturbations $\delta\varepsilon^{\mathrm{TT}}_{ij},\,\delta\mu^{\mathrm{TT}}_{ij}$ \\
    \bottomrule
  \end{tabular}

  \rowcolors{2}{}{}
\end{table}

Hence, this correspondence enables inverse design: starting from a chosen foliation or a target ADM evolution, one can translate it into conditions on a required space and time dependent profile of $\varepsilon_{ij}$ and $g_i$. This would enable engineers to use the rich library of general relativity solutions in their designs, which might be a powerful tool for transformation optics. Conversely, the same correspondence provides an electromagnetic interpretation of gravitational phenomena. In particular, we demonstrate the simplest case in general relativity, namely that linear perturbations of the mapped constraints in the TT gauge provide a toy model for cosmological gravitational waves in engineered media. This can potentially be extended to a wider class of cosmological and dynamical phenomena, and may serve as a useful benchmark for our theoretical understanding.

The dGREM formulation \cite{Olivares2022,Boyeneni2025} is a flux-conservative, first-order reformulation of the \((3{+}1)\) Einstein system, and it offers a natural companion framework for the field equation level mapping presented here. Thus, our theory can enrich Numerical Relativity and provide an alternative simulation framework, in which selected gravitational dynamics are realised through space and time-dependent electromagnetic constitutive parameters.

Ever since Einstein's breakthrough, mappings between the classical field theories of general relativity and Maxwell's electrodynamics have repeatedly resurfaced, of which the independent foundational contributions of Plebanski\cite{Plebanski1960} and Tamm\cite{Tamm1924} are prime examples. This persistent analogy continues to captivate the imaginations of successive generations of physicists. Ultimately, who can definitively say when an analogue system ceases to be a mere analogy and becomes an essential guide toward deeper layers of gravitation?

\textbf{Acknowledgements.}
E.E.E. thanks Adem Deniz Pişkin, Prof. Saul Teukolsky, and Prof. Kip Thorne for fruitful discussions on the ADM formalism, and Prof. Elias Most for introducing dGREM and whose thoughtful comments enriched the development of the ideas presented. Also, he is grateful to Prof. Mustafa Halilsoy and Prof. Mustafa Kuzuoğlu for their guidance. He also acknowledges the support of the Weizmann Institute of Science.
U.L. is supported by the Murray B. Koffler Professorial Chair.

\end{document}